\documentclass{article}
\usepackage{graphicx} 

\usepackage[english]{babel}
\usepackage{xcolor,xspace}
\usepackage{float}
\usepackage{listings}
\usepackage[letterpaper,top=2cm,bottom=2cm,left=3cm,right=3cm,marginparwidth=1.75cm]{geometry}

\usepackage{amsmath}
\usepackage{graphicx}
\usepackage[colorlinks=true, allcolors=black]{hyperref}
\usepackage[ruled,vlined,linesnumbered]{algorithm2e}
\usepackage[style=ieee]{biblatex} 
\newcommand{\agdes}{{\sc AGDES}\xspace}
\newtheorem{definition}{Definition}

\definecolor{codegray}{rgb}{0.95,0.95,0.95}
\definecolor{commentgray}{rgb}{0.5,0.5,0.5}
\definecolor{keywordblue}{rgb}{0.13,0.29,0.53}
\definecolor{stringgreen}{rgb}{0.11,0.62,0.11}

\title{\agdes: Automatic Generation of Dependent Event Sequences}
\author{Alexander Obeid Guzman \\
Univ. Grenoble Alpes, CNRS, Inria, Grenoble INP, LIG\\
\texttt{alexander.obeid-guzman@inria.fr}
}
\date{May 2026}

\begin{document}

\maketitle

\begin{abstract}
    This note presents \agdes, a tool for Automatic Generation of Dependent Event Sequences. Each event sequence is either generated as the output word of a deterministic finite automaton (DFA), or produced as the output word of a DFA called {\em transducer} that reads events from one or more input sequences, and produces an output sequence.
\end{abstract}

\section{Introduction}

Traditional causal discovery algorithms are well suited for continuous time-series data that can be i.i.d. sampled \cite{Gaoetal2024}. However, in complex dynamic systems such as telecommunication networks, critical data is not produced in a continuous way but it is rather discrete, asynchronous and event-driven \cite{networkHardware}. To develop causal discovery algorithms for these kind of systems there is a need for training and benchmarking data where the ground truth is known. In real world datasets most commonly the ground truth causal graph is unknown or heavily confounded by unmeasured variables and/or noise \cite{lackOfGT, lackOfGT2}. For these reasons we propose an automatic generation of dependent event sequences.

\section{Requirements}

\newcommand{\tuple}[1]{\langle #1 \rangle}

\agdes uses DFA to produce event sequences.

\begin{definition}[Deterministic finite automaton]
  A {\em deterministic finite automaton} (DFA) is a tuple $\tuple{Q,\Sigma,\rightarrow,q^0}$ where $Q$ is a finite set of states, $\Sigma$ is a finite alphabet, ${\rightarrow} \subseteq Q\times\Sigma\times Q$ is a transition relation, and $q^0\in Q$ is the initial state.
\end{definition}

\begin{definition}[Transducer]
  A transducer is a DFA $T = \tuple{Q,\Sigma,\rightarrow,q^0}$ with $Q = Q^i \uplus Q^o$, $\Sigma = \Sigma^i \uplus \Sigma^o$. $T$ is {\em deterministic} if each state having an outgoing transition labeled with an event in $\Sigma^o$, has no other outgoing transition.
  For a deterministic transducer $T = \tuple{Q,\Sigma,\rightarrow,q^0}$ with $\Sigma = \Sigma^i \uplus \Sigma^o$ and $x\in(\Sigma^i)^*$ let $T(x) = \{y\in(\Sigma^o)^* \mid \exists w\in L(T): \pi_{\Sigma^i}(w) = x \wedge \pi_{\Sigma^o}(w) = y\}$ where $\pi_k(w)$ is the {\em projection} of $w$ on $\Sigma_k$, that is, the word obtained by removing all events not in $\Sigma_k$ from $w$.
\end{definition}

We propose a highly configurable tool that fulfills the following requirements:
\begin{itemize}
  \item \textbf{Customization}: Alphabet size, number of states, ratio of input/output states, number of outgoing transitions per state should be user-defined parameters.
  \item \textbf{Noise control}: User can define level and type of noise. Noise can be introduced before producing the dependent sequences (resulting in noise propagation) or after the sequences are produced (representing noise in the observations). There should be three operations that introduce noise: removing, inserting and replacing characters. Note that for inserting and replacing, the character should come from the same alphabet $\Sigma$ to avoid conflicts with transducers reading the sequence.
  \item \textbf{Receptiveness of transducers}: All transducers in the network should be receptive which means that at all input states they should be able to react to any symbol from their input sequences.
  \item \textbf{Reachability}: All states in the producers and transducers generated should be reachable from the initial state and there should not be input-only cycles that prevent the generation of output symbols.
\end{itemize}

\section{Approach}
Our approach consists of generating a directed acyclic graph (DAG) of transducers at internal nodes and producers at source nodes. In the simplest case, we can take a DAG where $X \longrightarrow Y$ ($X$ causes $Y$) and represent this with just one producer $P$ and one transducer $T$. $P$ and $T$ are generated randomly.\\

We can control different parameters during the generation of the $P$ automaton:
\begin{itemize}
    \item Number of states (min, max, skewness)
    \item Alphabet size (min, max, skewness)
    \item Transitions per state (min, max, skewness)
\end{itemize}
By default, the parameters are sampled from a skew-normal distribution \href{https://docs.scipy.org/doc/scipy/reference/generated/scipy.stats.skewnorm.html}{\textcolor{blue}{skewnorm}} but the user can also specify other distributions from \href{https://docs.scipy.org/doc/scipy/reference/stats.html}{\textcolor{blue}{scipy.stats}} if it better suits their needs.\\

For the $T$ transducer we can control the same parameters as for the producer in addition to:
\begin{itemize}
    \item Ratio between input/output states (approximate). This ratio helps control the amount of symbols that will be read from the input.
\end{itemize}

An example of a producer $P$ generated with this method is:
\begin{figure}[h!]
    \centering
    \includegraphics[width=0.5\linewidth]{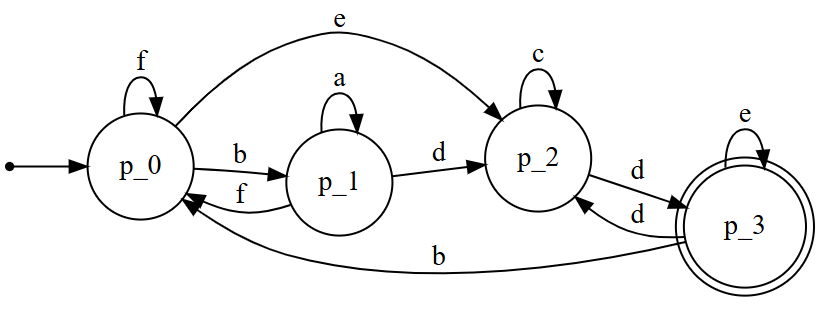}
    \caption{Producer automaton $P$ generated with minStates=4, maxStates=6, minAlphabet=3, maxAlphabet=10, minTransitions=1, maxTransitions=4}
    \label{fig:producer}
\end{figure}

And an example of a transducer is the following:
\begin{figure}[H]
    \centering
    \includegraphics[width=0.9\linewidth]{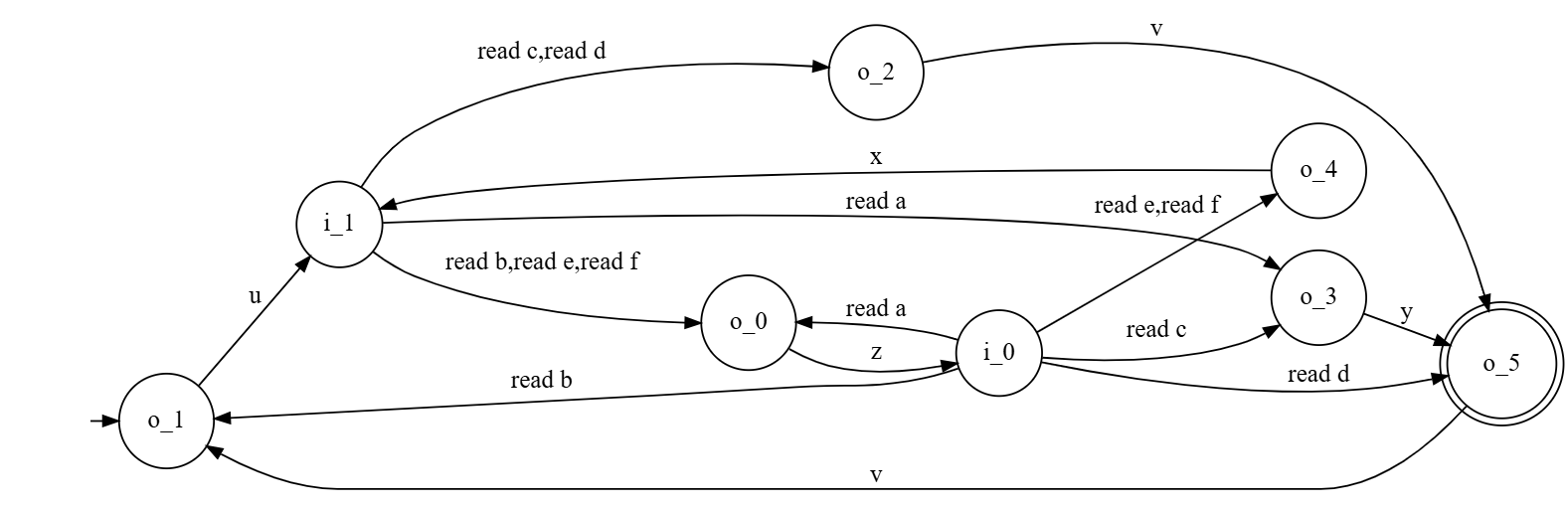}
    \caption{Transducer automaton $T$ generated with minStates=4, maxStates=10, minAlphabet=3, maxAlphabet=10, minTransitions=1, maxTransitions=1, ratio\_i\_o=0.3}
    \label{fig:transducer}
\end{figure}

Now that we have a producer $P$, we can generate random words of length $k$ from $P$ by just randomly taking one of the possible transitions from the current state (starting from the initial state). For example, in Figure \ref{fig:producer}, from the initial state $p_0$ we can choose between $\{f,b,e\}$, if we choose $b$ we are now in state $p_1$ and can now produce $\{a, d, f\}$ and so on. A valid sequence of length 10 produced with $P$ would be $x=\text{bfbfedeeee}$.
With $T$ and $x$ we can now produce a sequence $y$. To do so, we start from the initial state of $T$, in this case, $o_1$ and as we only have one possible transition, we will generate the first character of $y$ which will be a ``u". Then, we are in state $i_1$ which is an input state, so we will read the next available character from $x$, which in this case is a ``b" which will take us to $o_0$ from which we will produce a ``z" and go to $i_0$. Following this logic, if we want to produce $y$ of length 10 with $T(x=\text{bfbfedeeee})$ resulting in $y=\text{uzxzxzvuzx}$. With this methodology, by construction, we introduce the causal direction $x\longrightarrow y$

\section{Using \agdes}

You can find \agdes in the following repository \href{https://gitlab.inria.fr/aobeidgu/agdes}{\textcolor{blue}{\agdes}} \cite{AGDES}

\subsection{Requirements}
To run \agdes, ensure that your environment meets the following prerequisites:

\begin{itemize}
    \item \textbf{Python}: Version 3.10 or higher
    \item \textbf{Dependencies}:
    \begin{itemize}
        \item numpy
        \item pandas
        \item scipy.stats
        \item automata-lib
        \item graphviz
    \end{itemize}
\end{itemize}

\subsection{Available functionality}
\subsubsection{generate\_random\_producer}
\begin{verbatim}
(minStates=1, maxStates=6, skw_states=0, dist_states=None, 
 minAlphabet=1, maxAlphabet=8, skw_alphabet=0, dist_alphabet=None, 
 minTransitions=1, maxTransitions=4, skw_transitions=0, dist_transitions=None, 
 symbol='q_', verbose=False)
\end{verbatim}
Generates a random producer (DFA) with customizable probability distributions for its states, alphabet size, and transitions.\\

\textbf{Parameters:}
\begin{description}
    \item[\texttt{minStates} (\textit{int})] Minimum number of states to generate. [1, 30] default is 1. 
    \item[\texttt{maxStates} (\textit{int})] Maximum number of states to generate. [minStates, 50] default is 6.
    \item[\texttt{skw\_states} (\textit{float})] Skewness parameter for the state generation skewnorm distribution [$-\infty$, $\infty$] default is 0.
    \item[\texttt{dist\_states} (\textit{object})] A frozen \texttt{scipy.stats} distribution object for sampling the number of states. Default is \texttt{None}. If specified, skw\_states is ignored.
    
    \item[\texttt{minAlphabet} (\textit{int})] Minimum number of symbols in the generated alphabet. [1, 26] default is 1. Symbols are lower case letters from the Latin alphabet (can be generalized).
    \item[\texttt{maxAlphabet} (\textit{int})] Maximum number of symbols in the generated alphabet. [minAlphabet, 26] default is 8. Symbols are lower case letters from the Latin alphabet (can be generalized).
    \item[\texttt{skw\_alphabet} (\textit{float})] Skewness parameter for the alphabet size skewnorm distribution. [$-\infty$, $\infty$] default is 0.
    \item[\texttt{dist\_alphabet} (\textit{object})] A frozen \texttt{scipy.stats} distribution object for sampling the number of states. Default is \texttt{None}. If specified, skw\_alphabet is ignored.
    
    \item[\texttt{minTransitions} (\textit{int})] Minimum number of outgoing transitions per state. [1, maxAlphabet] default is 1.
    \item[\texttt{maxTransitions} (\textit{int})] Maximum number of outgoing transitions per state. [1, maxAlphabet] default is 4.
    \item[\texttt{skw\_transitions} (\textit{float})] Skewness parameter for the transition count skewnorm distribution. [$-\infty$, $\infty$] default is 0.
    \item[\texttt{dist\_transitions} (\textit{object})] A frozen \texttt{scipy.stats} distribution object for sampling the number of states. Default is \texttt{None}. If specified, skw\_transitions is ignored.
    
    \item[\texttt{symbol} (\textit{str})] The prefix string used for naming the generated states (e.g., resulting in \texttt{q\_0}, \texttt{q\_1}). Default is \texttt{'q\_'}.
    \item[\texttt{verbose} (\textit{bool})] If \texttt{True}, outputs detailed generation logs to the console. Default is \texttt{False}.
\end{description}
\vspace{0.3cm}
\noindent\textbf{Returns:}
\begin{description}
    \item[\textit{DFA}] A random producer object generated according to the specified constraints and distributions.
\end{description}
\subsubsection{generate\_random\_transducer}
\begin{verbatim}
(read_input_alphabet, ratio_i_o=0.3, minStates=1, maxStates=6, 
 skw_states=0, dist_states=None, minAlphabet=1, maxAlphabet=8, 
 skw_alphabet=0, dist_alphabet=None, minTransitions=1, maxTransitions=1, 
 skw_transitions=0, dist_transitions=None, verbose=False)
\end{verbatim}
Generates a random finite state transducer based on a provided input alphabet and customizable probability distributions for its states, output alphabet, and transitions.\\

\textbf{Parameters:}
\begin{description}
    \item[\texttt{read\_input\_alphabet} (\textit{set})] The set of input symbols the transducer is capable of reading.
    \item[\texttt{ratio\_i\_o} (\textit{float})] The target approximated ratio of input to output states. [0, 1] default is 0.3.
    \item[\texttt{minStates} (\textit{int})] Minimum number of states to generate. [1, 30] default is 1. 
    \item[\texttt{maxStates} (\textit{int})] Maximum number of states to generate. [minStates, 50] default is 6.
    \item[\texttt{skw\_states} (\textit{float})] Skewness parameter for the state generation skewnorm distribution [$-\infty$, $\infty$] default is 0.
    \item[\texttt{dist\_states} (\textit{object})] A frozen \texttt{scipy.stats} distribution object for sampling the number of states. Default is \texttt{None}. If specified, skw\_states is ignored.
    
    \item[\texttt{minAlphabet} (\textit{int})] Minimum number of symbols in the generated alphabet. [1, 26] default is 1. Symbols are lower case letters from the Latin alphabet (can be generalized).
    \item[\texttt{maxAlphabet} (\textit{int})] Maximum number of symbols in the generated alphabet. [minAlphabet, 26] default is 8. Symbols are upper case letters from the Latin alphabet (can be generalized).
    \item[\texttt{skw\_alphabet} (\textit{float})] Skewness parameter for the alphabet size skewnorm distribution. [$-\infty$, $\infty$] default is 0.
    \item[\texttt{dist\_alphabet} (\textit{object})] A frozen \texttt{scipy.stats} distribution object for sampling the number of states. Default is \texttt{None}. If specified, skw\_alphabet is ignored.
    
    \item[\texttt{minTransitions} (\textit{int})] Minimum number of outgoing transitions per state. [1, maxAlphabet] default is 1.
    \item[\texttt{maxTransitions} (\textit{int})] Maximum number of outgoing transitions per state. [1, maxAlphabet] default is 1 (this makes it a deterministic transducer).
    \item[\texttt{skw\_transitions} (\textit{float})] Skewness parameter for the transition count skewnorm distribution. [$-\infty$, $\infty$] default is 0.
    \item[\texttt{dist\_transitions} (\textit{object})] A frozen \texttt{scipy.stats} distribution object for sampling the number of states. Default is \texttt{None}. If specified, skw\_transitions is ignored.
    \texttt{q\_1}). Default is \texttt{'q\_'}.
    \item[\texttt{verbose} (\textit{bool})] If \texttt{True}, outputs detailed generation logs to the console. Default is \texttt{False}.
\end{description}
\vspace{0.3cm}
\noindent\textbf{Returns:}
\begin{description}
    \item[\textit{DFA}] A random transducer object configured with the specified input alphabet, states, and transitions constraints.
\end{description}

\subsubsection{random\_word\_from\_producer}
\begin{verbatim}
(producer, output_length=10)
\end{verbatim}
Generates a random word (sequence of symbols) of a specified length by traversing the provided producer automaton.\\

\textbf{Parameters:}
\begin{description}
    \item[\texttt{producer P} (\textit{DFA})] The producer DFA object used to generate the random word.
    \item[\texttt{output\_length} (\textit{int})] The desired length of the generated sequence. [1, $\infty$] default is 10.
\end{description}

\vspace{0.3cm}
\noindent\textbf{Returns:}
\begin{description}
    \item[\textit{str}] The randomly generated sequence of symbols $w \in L(P)$ of length \texttt{output\_length}.
\end{description}

\subsubsection{random\_word\_from\_transducer}
\begin{verbatim}
(transducer, input_word, output_length=10, return_order=False)
\end{verbatim}

\noindent Generates an output word by processing a given input sequence through the provided finite state transducer.\\

\textbf{Parameters:}
\begin{description}
    \item[\texttt{transducer T} (\textit{DFA})] The transducer object used to read the input and generate the corresponding output sequence.
    \item[\texttt{input\_word x} (\textit{str})] The sequence of input symbols to be fed into the transducer.
    \item[\texttt{output\_length} (\textit{int})] The desired length constraint for the generated output sequence. [1, $\infty$] default is 10.
    \item[\texttt{return\_order} (\textit{bool})] If \texttt{True}, includes the interleaving between input and ouput symbols traversed by the transducer. Default is \texttt{False}.
\end{description}

\vspace{0.3cm}
\noindent\textbf{Returns:}
\begin{description}
    \item[\textit{str}, or \textit{tuple}] The generated output sequence $y=T(x)$ resulting from the input word $x$. If \texttt{return\_order} is \texttt{True}, it returns a tuple containing the output word alongside its associated interleaving with input symbols.
\end{description}

\textbf{Warning:} Depending on the transducer, it may happen that in order to produce an output word $y$ of length $k$ for an input word $x$, $|x| > k$.

\subsubsection{introduce\_insert\_or\_delete\_noise}
\begin{verbatim}
(word, n_symbols_change=None, noise_level=0.1, prob_insert=0.5)
\end{verbatim}
Introduces noise into a given word by randomly inserting or deleting symbols from the same alphabet based on specified probabilities and noise levels. The output word may be of different size than the input word.\\

\textbf{Parameters:}
\begin{description}
    \item[\texttt{word} (\textit{str})] The original sequence of symbols to be modified.
    \item[\texttt{n\_symbols\_change} (\textit{int}, optional)] The exact number of symbols to insert or delete [0, $|word|$]. If set to \texttt{None}, the number of changes is calculated dynamically using the \texttt{noise\_level}. Default is \texttt{None}. If specified, \texttt{noise\_level} is ignored.
    \item[\texttt{noise\_level} (\textit{float})] The proportion of the word's length to alter when \texttt{n\_symbols\_change} is not provided. [0, 1] default is 0.1 (representing 10\% noise).
    \item[\texttt{prob\_insert} (\textit{float})] The probability [0, 1] that a noise operation will be an insertion. The probability of a deletion is automatically $1 - \texttt{prob\_insert}$. Default is 0.5.
\end{description}

\vspace{0.3cm}
\noindent\textbf{Returns:}
\begin{description}
    \item[\textit{str}] The modified, noisy word after the insertion and deletion operations have been applied. Note that a replacement is equivalent to a deletion and an insertion.
\end{description}

\subsubsection{introduce\_replacement\_noise}
\begin{verbatim}
(word, n_symbols_change=None, noise_level=0.1)
\end{verbatim}

\noindent Introduces noise into a given word by randomly replacing existing symbols with new ones, ensuring that the overall length of the word remains unchanged.\\

\textbf{Parameters:}
\begin{description}
    \item[\texttt{word} (\textit{str})] The original sequence of symbols to be modified.
    \item[\texttt{n\_symbols\_change} (\textit{int}, optional)] The exact number of replacements to do [0, $|word|$]. If set to \texttt{None}, the number of changes is calculated dynamically using the \texttt{noise\_level}. Default is \texttt{None}. If specified, \texttt{noise\_level} is ignored.
    \item[\texttt{noise\_level} (\textit{float})] The proportion of the word's length to alter when \texttt{n\_symbols\_change} is not provided. [0, 1] default is 0.1 (representing 10\% noise).
\end{description}

\vspace{0.3cm}
\noindent\textbf{Returns:}
\begin{description}
    \item[\textit{str}] The modified, noisy word containing the replaced symbols, retaining its exact original length.
\end{description}

\section{Examples} 
A basic usage example will be given here. For further examples and higher customization look at the Examples.ipynb notebook in \href{https://gitlab.inria.fr/aobeidgu/agdes}{\textcolor{blue}{\agdes}}.\\

To generate a producer one simply calls the generate\_random\_producer function with the desired parameters.
\begin{lstlisting}[language=Python]
producer = generate_random_producer(minStates=3, maxStates=6, skw_states=4, minAlphabet=3, maxAlphabet=8, skw_alphabet=0, minTransitions=1, maxTransitions=4, skw_transitions=0, symbol='q_', verbose=False)
\end{lstlisting}

To generate a transducer one calls the generate\_random\_transducer function with the desired parameters and passes the input\_symbols attribute from the producer (its alphabet) in the read\_input\_alphabet parameter.
\begin{lstlisting}[language=Python]
p_alpha = producer.input_symbols
transducer = generate_random_transducer(read_input_alphabet=p_alpha, ratio_i_o=0.3, minStates=3, maxStates=6, skw_states=4, minAlphabet=3, maxAlphabet=8, skw_alphabet=0, minTransitions=1, maxTransitions=4, skw_transitions=0, verbose=False)
\end{lstlisting}
Now that we have a producer and a transducer, we can model the causal DAG $X \longrightarrow Y$ by generating strings $x$ from the producer and feeding it to the transducer to produce $y$. An example of how to do this is shown in the following lines.
\begin{lstlisting}[language=Python]
x = random_word_from_producer(producer=producer, output_length=10)
y = random_word_from_transducer(transducer=transducer, input_word=x, output_length=10)
\end{lstlisting}
By generating sequences like this, we know that $x$ is causing $y$ but not the other way around. \\

To introduce noise, we can use the functions introduce\_insert\_or\_delete\_noise and introduce\_replacement\_noise. Note that the first one may produce a noisy word of different length than the original and the latter ensures the noisy word has the same length as the original one. 
\begin{lstlisting}[language=Python]
noisy_x = introduce_insert_or_delete_noise(word=x, noise_level=0.2, prob_insert=0.5)
noisy_y = introduce_replacement_noise(word=y, noise_level=0.2)
\end{lstlisting}
This kind of noise is noise in the observations of $x$ and $y$ but there is no propagation of noise from $x$ to $y$ in this setting. As both functions insert noise from the same alphabet, it is possible to feed a noisy sequence from the producer to a transducer resulting in the noise propagation from $x$ to $y$. 
\begin{lstlisting}[language=Python]
y_with_noisy_x = random_word_from_transducer(transducer=transducer, input_word=noisy_x, output_length=10)
\end{lstlisting}
This is different than independently inserting noise into both sequences after the production of $y$.

\section{Discussion}
\agdes sets the basics for dependent sequence events generation. It allows a high customization of the DFA used for the production of the sequences as well as high control on the noise. Currently, it supports linear DAGs (or chains) of the type $X_1\longrightarrow X_2\longrightarrow \dots \longrightarrow X_n$. In the future we plan to extend this to allow each transducer to read from multiple inputs at the same time. We also plan on supporting the automatic generation of DAGs given usual network metrics like distribution of connectedness. \\

Another area of improvement is in the way we control the dependency between the generated sequences. At the moment, we can only control the ratio between input and output states of a transducer. This indirectly controls the amount of input information used by the transducer to produce the output, but it is not directly controlling the dependency level between the input and output sequences. Also, as we are using DFA to produce the sequences, in a transducer, given a state and an input symbol, the output symbol will always be the same. One alternative to introduce non-determinsim (and maybe a way of controlling the dependency between sequences) is that instead of using DFA we use Markov Decision Processes (MDP). By doing so, we can introduce probabilities of transitions, so from a given state and input symbol we may have a probability of using the symbol to determine the output or not using the symbol at all. 
\printbibliography

\end{document}